# SpineMamba: Enhancing 3D Spinal Segmentation in Clinical Imaging through Residual Visual Mamba Layers and Shape Priors

Zhiqing Zhang[a,b], Tianyong Liu[c], Guojia Fan[d], Bin Li[a], Qianjin Feng[f,*] and Shoujun Zhou[a,**]

[a]*Shenzhen Institute of Advanced Technology, Chinese Academy of Sciences, Shenzhen, 518055, China*

[b]*University of Chinese Academy of Sciences, Street 29, 100049 Beijing, China*

[c]*Institute of Unconventional Oil & Gas Research,Northeast Petroleum University., Street 15, Daqing, 163318, China*

[d]*College Of Information Science and Engineering, Northeastern University, liaoning, 110819, , china*

[f]*School of Biomedical Engineering, Southern Medical University, Guangzhou, 510515, China*

## ABSTRACT

Accurate segmentation of 3D clinical medical images is critical in the diagnosis and treatment of spinal diseases. However, the inherent complexity of spinal anatomy and uncertainty inherent in current imaging technologies, poses significant challenges for semantic segmentation of spinal images. Although convolutional neural networks (CNNs) and Transformer-based models have made some progress in spinal segmentation, their limitations in handling long-range dependencies hinder further improvements in segmentation accuracy. To address these challenges, we introduce a residual visual Mamba layer to effectively capture and model the deep semantic features and long-range spatial dependencies of 3D spinal data. To further enhance the structural semantic understanding of the vertebrae, we also propose a novel spinal shape prior module that captures specific anatomical information of the spine from medical images, significantly enhancing the model's ability to extract structural semantic information of the vertebrae. Comparative and ablation experiments on two datasets demonstrate that SpineMamba outperforms existing state-of-the-art models. On the CT dataset, the average Dice similarity coefficient for segmentation reaches as high as 94.40±4%, while on the MR dataset, it reaches 86.95±10%. Notably, compared to the renowned nnU-Net, SpineMamba achieves superior segmentation performance, exceeding it by up to 2 percentage points. This underscores its accuracy, robustness, and excellent generalization capabilities

## 1. Introduction

The spine is the body's second lifeline, supporting the normal functioning of various organs. The central nervous system within the spine is intricately connected throughout the body's network of meridians, serving as the core of the human body's neural pathways. Any damage to the spine can affect the transmission and functioning of nerves, potentially leading to diseases in related tissues. The evolution of modern digital orthopedics hinges on the segmentation of 3D medical imaging data, as illustrated in Figure 1. This technology has multiple applications in the clinical treatment of orthopedic diseases, including diagnosis and treatment [1], preoperative planning [2], and real-time image navigation during surgery [3, 4]. The increasing demand for spinal clinical imaging has intensified the burden on physicians, making manual annotation laborious and time-consuming. In contrast, automated annotation is more efficient and less prone to subjective influence. However, accurately segmenting the spine in volumetric images poses several challenges. First, inherent characteristics such as high noise and low contrast in different 3D medical imaging modalities can cause artifacts in 3D CT scans, leading to blurred and distorted spinal images. Similarly, anisotropic spatial resolution in spinal MR images results in intensity inhomogeneity (Figure 2(b)) and partial volume effects (Figure 2(c)), contributing to data heterogeneity [5] and complicating the segmentation process. Second, spinal MR images exhibit inter-class similarities and intra-class variations [6]. As shown in Figures 2(c) and (d), there are inter-class similarities within a single sample and across different samples, while intra-class variations arise in samples with or without lumbarization. Third, the high computational memory demands associated with high-dimensional images pose significant challenges to algorithmic models.

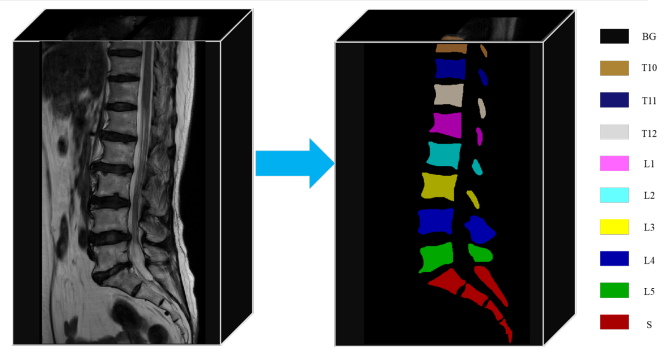

**Fig. 1**: Schematic diagram of vertebral segmentation in 3D MRI images. This image contains 10 labels. "BG" stands for background. "T," "L," and "S" represent thoracic vertebrae, lumbar vertebrae, and sacrum, respectively.

With the rapid development of deep learning (DL) technology, DL-based segmentation methods have demonstrated effectiveness in delineating organs/tumors and reducing labor costs. Currently, existing DL-based segmentation methods include convolutional neural network (CNN)-based approaches [7, 8],and Transformer-based variants [9, 10]. Classic CNN-based models, such as UNet [7], have been widely adopted for segmentation tasks involving medical organs and lesions. UNet's symmetric structure and skip connections enable efficient segmentation performance. However, the effectiveness of UNet is constrained by the inherent reliance on convolution operations within the CNN architecture, which limits the model's ability to capture long-range spatial dependencies between pixels, thereby restricting its capacity to extract global features. Some studies have attempted to address the inherent locality of convolution operations by incorporating additional convolutional layers [11] or self-attention mechanisms [12], but limitations persist in modeling long-range dependencies. In contrast to CNN architectures, Transformers [13] excel at capturing global information. Transformer-based architectures do not emphasize the spatial hierarchy of images but instead treat images as sequences of continuous patches. For instance, Swin-Unet [14], which combines the Swin Transformer [15] with a U-shaped architecture, and TransUnet [9], which incorporates ViT [16] for feature extraction during the encoding phase and utilizes convolutional kernels in the decoding phase, have improved the models' ability to handle long-range dependencies. However, the computational cost of self-attention mechanisms increases quadratically with input size, introducing greater computational complexity [16, 13, 17]. Overall, although these traditional deep learning methods exhibit strong performance in certain aspects, they still face limitations in handling long-range dependencies and computational complexity. In spinal segmentation tasks, effectively capturing long-range dependencies between vertebrae is crucial for accurate segmentation and avoiding semantic confusion between vertebrae. Therefore, developing a novel medical image segmentation architecture that can capture strong long-range information between spinal vertebrae while maintaining linear computational complexity is an urgent issue that needs to be addressed.

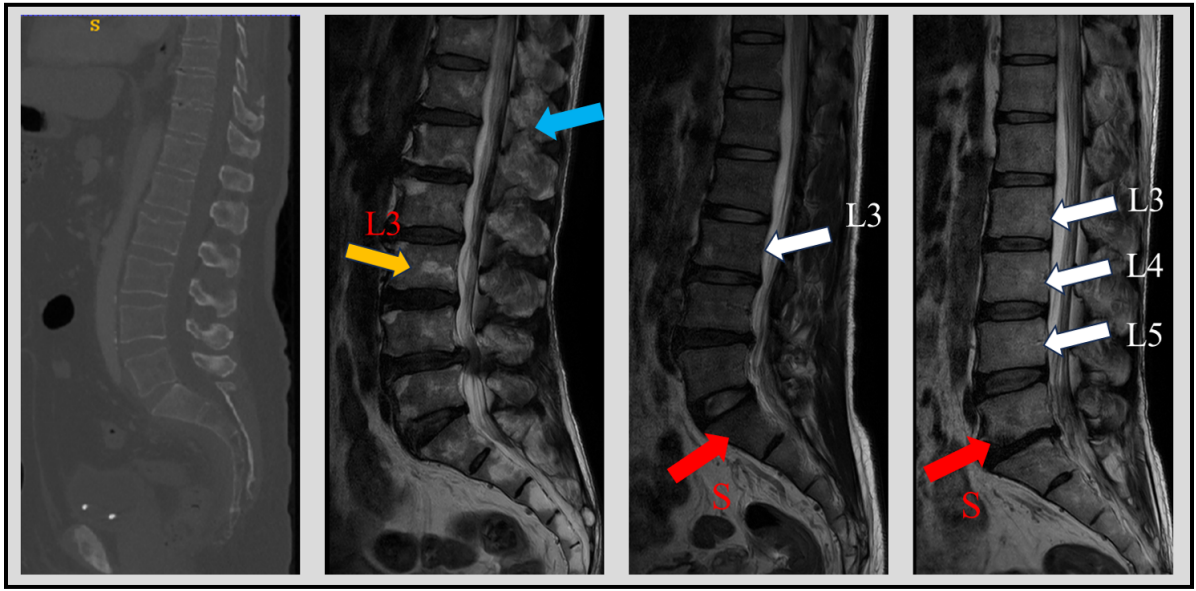

**Fig. 2:** Challenges in spinal segmentation in 3D images.(a) Blurred boundaries caused by artifacts in CT imaging.(b) Intensity inhomogeneity between the edge and center of the L3 vertebral body in MR imaging (indicated by the yellow arrow). Partial volume effects result in blurred pedicle edges (indicated by the blue arrow).(c) In spinal MR images, there is inter-class similarity within the same sample (e.g., L3, L4, and L5 in (d), indicated by white arrows) and between different samples (e.g., L3 in (c) and L4 in (d), indicated by white arrows). Additionally, differences in the appearance of sacral vertebrae are observed between different samples (indicated by red arrows in (c) and (d)). The sacrum in (c) does not exhibit lumbarization, whereas lumbarization is observed in (d).

Recently, state-space models (SSMs) [18], particularly Structured State Space Sequence Models (S4) [18], have emerged as efficient and effective building blocks (or layers) for constructing deep networks, achieving state-of-the-art performance in the analysis of continuous long-sequence data. Mamba [19] further improves S4 with a selection mechanism, capturing long-range interactions while maintaining linear computational complexity. Inspired by the success of Mamba in medical image segmentation tasks [20, 21, 22] , this study proposes a Mamba-based spinal segmentation network architecture, SpineMamba, to address existing challenges. By leveraging the strengths of state-space models, we designed a novel U-shaped network structure specifically for the segmentation of 3D spinal medical images. In particular, we developed a learnable shape prior module within the Mamba architecture that can be embedded into the U-shaped network to guide precise segmentation, effectively addressing the limitations of current deep learning networks in capturing spinal morphological priors.

The main contributions of this paper are as follows:

1. The first Mamba-based 3D spinal segmentation model: This model maintains linear computational complexity while capturing both local fine-grained information and long-range relationships inside pictures through the hybridization of SSM and CNN architectures.

2. The creative design of a learnable 3D spinal shape prior module: This module effectively integrates as a plug-and-play part of the network, learning spinal priors during training to improve the predicted spinal region boundaries and get past issues with intra- and inter-class variability. As a result, spinal segmentation performance is significantly improved.

3. Excellent segmentation performance and strong generalization: The suggested method outperforms a number of cutting-edge methods on two separate datasets with varying modalities, exhibiting strong generalizability and robustness and offering insightful information for the creation of more effective and efficient SSM-based spinal segmentation methods.

## 2. Related work

### 2.1. vertebral segmentation

At the core of the spinal segmentation workflow is image segmentation, which involves assigning specific labels to each vertebral voxel (the smallest unit of the image) in a medical image to generate accurate segmentation masks. This process not only provides a detailed map of each spinal structure within the image but also ensures that each label corresponds to a unique instance of a vertebra. These generated segmentation masks form the foundation for the automatic quantification of spinal biomarkers, playing a crucial role in clinical applications such as supporting disease diagnosis and treatment by detecting and localizing pathological changes and abnormalities.

Early spinal segmentation primarily relied on traditional modeling methods, such as mathematical models and hand-crafted features, including deformable models [23, 24], custom filters [25], atlas-based methods [26, 27], and machine learning approaches [28, 29]. These methods were widely used in spinal segmentation research. However, these traditional methods struggled to handle inter-individual variability and had limited modeling capabilities, requiring manual feature extraction, which restricted segmentation performance.

With the advancement of deep learning technology, several studies on spinal segmentation have successfully applied neural network-based models [30, 31]. These methods can be subdivided into two major categories based on the structural approach to segmentation . The first category involves separately segmenting vertebrae (VB) and intervertebral discs (IVD), focusing on accurately distinguishing and quantifying each structure. For instance, Tao et al. [17] proposed a Transformer-based method to address VB labeling and segmentation, further enhancing performance through multitask learning and achieving unique predictions by designing a global loss function and a lightweight Transformer architecture. Zhang et al. [32] introduced a Sequential Conditional Reinforcement Learning Network (SCRL) to address the simultaneous detection and segmentation of VBs in MR spinal images, achieving accurate detection and segmentation results for spinal diseases. In terms of IVD segmentation, Li et al. [33] proposed an innovative three-dimensional multi-scale contextual fully convolutional network specifically designed for localizing and segmenting intervertebral discs from multimodal 3D MR data. This network effectively integrates information from different modalities through an advanced voxel-selective dropout strategy, significantly enhancing the network's ability to learn and recognize complex structures.

The second category involves the simultaneous segmentation of vertebrae and intervertebral discs, aiming to improve segmentation efficiency while retaining critical diagnostic information, thereby ensuring or enhancing the accuracy of spinal disease segmentation. This simultaneous segmentation task is more challenging, requiring the model to not only accurately distinguish between the two structures but also capture the complex spatial relationships between them. Algorithms need to possess high-resolution image processing capabilities and sensitivity to subtle structural differences to achieve high-quality segmentation results. Han et al. [1] developed Spine-GAN, a 2D semantic segmentation network for multiple spinal structures, capable of segmenting and classifying six types of spinal structures in a single step. It consists of a segmentation network and a discriminator network, achieving high pixel-level segmentation accuracy. However, it overlooks anatomical priors of spinal structures and loses the spatial correlation of anatomical regions. To reduce inter-class similarity in spinal MR images, Pang et al. [6] first performed coarse segmentation on 3D MR images, followed by 2D fine segmentation, thereby enhancing the semantic information of spinal images within the segmentation network. Nevertheless, this method has some minor shortcomings in handling data imbalance and boundary blurring issues.

### 2.2. Shape Priors

In the field of medical image segmentation, the application of shape knowledge has become a critical factor in improving segmentation accuracy and efficiency. Numerous studies have attempted to incorporate shape priors into the design of segmentation models, as shape knowledge mimics the expertise of clinicians, enabling more accurate and consistent use of anatomical shape information. Shape priors can be broadly divided into two categories based on the design approach: traditional methods and deep learning-based shape prior models. Traditional methods include atlas-based models [34] and statistical models [35].

Atlas-based methods use pre-labeled images (atlases) as references. The target image is aligned with the atlas, and segmentation is conducted by transferring the atlas labels.

This method benefits significantly from the introduction of shape knowledge, as the atlas provides detailed anatomical references, improving the segmentation accuracy of complex structures. However, the performance of these methods also depends on the quality and quantity of the atlas and its alignment with new images. The registration process can be time-consuming, and selecting the source images for registration can be challenging [36, 37]. Statistical models, on the other hand, utilize shape information learned from a set of training images to assist in the recognition and segmentation of structures in new images. Representative examples include Active Shape Models (ASMs) [38] and Active Appearance Models (AAMs) [39]. When appropriately trained, these models can generalize to new data from different imaging modalities, addressing anatomical variations between individuals. However, the performance of these methods is highly dependent on the quality and diversity of the training data. If the training set is not comprehensive or is biased, the model may fail to accurately generalize to new data, requiring regular updates and maintenance to retain accuracy and relevance.

In recent years, the rise of deep learning technologies has brought revolutionary changes to the field of medical image segmentation. Particularly, the introduction of Convolutional Neural Networks (CNNs) has become the core of contemporary segmentation technology due to their ability to extract complex hierarchical features from large datasets. Among the many CNN architectures, U-Net and its derivative models have gained considerable attention for their outstanding performance in medical image segmentation tasks. These models not only enhance segmentation accuracy but also optimize processing speed, significantly advancing the development of medical image analysis technology. U-shaped networks [7] can automatically extract multi-scale features, including semantic and detailed information, from specific regions in the encoder. These features are then combined with deep features from the bottleneck via skip connections in the decoder structure, achieving excellent segmentation performance with a streamlined architecture. However, current U-shaped network-based models also have limitations, such as the inability to utilize specific anatomical or shape knowledge [40] and challenges in learning inductive biases [41].

To address these shortcomings in U-shaped networks, previous works, such as deformable convolutions [42], have focused on this direction. Deformable convolutions enhance the representation of specific region shapes by introducing shape characteristics to the convolutional kernels. Jurdi et al. [43] proposed BB-UNet, a deep learning model that integrates location and shape priors before model training and merges them into the skip connections through novel convolutional layers. The proposed architecture helps direct the attention kernels in neural training to guide the model in locating organs. Additionally, it fine-tunes the encoder layers based on location constraints. Nguyen et al. [44] introduced the Cascaded Context Module (CCM) and the Balanced Attention Module (BAM), which implement attention mechanisms in background, speckle, and boundary regions, respectively, thereby highlighting the contextual features of polyps. Although the performance of existing models can be enhanced by incorporating implicit shape priors, they often lack sufficient interpretability and generalizability when dealing with organs of different morphologies.

Unlike implicit shape priors, explicit shape priors incorporate shape information directly into the model as input. This approach allows for a more explicit consideration of the shape and structural characteristics of the target object, enabling more effective interpretation and adjustment during training. In addition, explicit priors are often learnable, which equips the model with the ability to continuously self-optimize throughout the training process. This capability not only gradually enhances the model's performance but also significantly improves its robustness across different modalities. For example, Meng et al. [45] developed DedustGAN, which incorporates a learning mechanism during image generation. The adaptability and practical utility of the model are greatly improved by this creative design. In a similar vein, You et al. [46]'s shape module has proven to be exceptionally successful at processing multimodal images from three distinct anatomical locations. Explicit shape priors so aid in making better use of this data to direct the model throughout the segmentation of the target item. Therefore, developing an explicit shape prior method would improve the model's performance in a variety of medical picture segmentation tasks by strengthening its capacity to differentiate features across various areas.

## 2.3. Mamba Model Based on SSM

In recent advancements in deep learning, Convolutional Neural Networks (CNNs) [7, 8] and Vision Transformers (ViTs) [10, 47, 48] have become key benchmarks for medical image segmentation tasks. CNNs typically employ an encoder-decoder structure [30, 31, 49] with skip connections, enabling the decoder to reuse features extracted by the encoder. However, CNNs are primarily adept at capturing local features and struggle to effectively capture long-range spatial correlations between pixels, even when these local regions exist within the broader context. This limitation hinders their ability to extract global features. In contrast, the Transformer architecture, leveraging self-attention mechanisms [50], has significantly improved the understanding of global context, leading to breakthroughs in global feature modeling. The adoption of innovative techniques such as shifted windows has further advanced ViT (Vision Transformer) and its derivative architectures like the Swin Transformer [15]. Additionally, integrating ViTs with CNN architectures, as seen in models like TransUNet [9] and UNETR [51], has expanded the applicability of Transformers in computer vision, enhancing their flexibility and effectiveness in handling visual tasks. Despite their success in capturing long-range dependencies, Transformers face computational efficiency challenges due to the quadratic growth of the self-attention mechanism's computational cost with input size [17, 52].

Recently, progress in state-space models (SSMs) has garnered attention, particularly the structured state-space model (S4) [18], which demonstrates high efficiency in processing long-sequence data, marking it as a promising solution [53, 54]. The Visual Mamba model builds on these advancements, as illustrated in Figure 3, introducing innovations such as the Cross-Scan Module (CSM) [19, 55] and integrated optimization of convolution operations. The CSM performs an orderly traversal within the spatial domain, converting non-causal visual images into an ordered sequence of patches. Within the Visual State Space (VSS) block, the input features first undergo processing through a linear embedding layer and are then split into two paths. One path undergoes depthwise convolution [56] and SiLU activation [57], followed by processing in the SS2D module and layer normalization, before merging with the other path, which also passes through SiLU activation. Unlike traditional vision Transformers, the VSS block omits positional embeddings and adopts a simplified structure without MLP (Multi-Layer Perceptron), enabling more compact block stacking at the same network depth, thereby improving the model's efficiency and performance. Specifically, SegMamba [20] ingeniously integrates SSM into the encoder stage while retaining traditional Convolutional Neural Networks (CNNs) in the decoder stage, creating an SSM-CNN hybrid model for 3D brain tumor segmentation tasks. This design not only leverages the advantages of SSMs in capturing global features but also preserves the efficiency of CNNs in handling local details. U-Mamba [21] further innovates by proposing a novel SSM-CNN hybrid model that features linear scaling of feature dimensions. This capability allows the model to simultaneously capture local fine-grained features and long-range dependencies in images, resulting in significant improvements in segmentation accuracy and robustness. VM-UNet [22] is the first to explore the potential application of a pure SSM model in the field of medical image segmentation, achieving remarkable success across three datasets, demonstrating the potential and value of SSMs in medical image analysis.

## 3. Methodology

### 3.1. Framework overview

Inspired by these advantages, we propose incorporating Visual Mamba blocks (VSS) within the U-shaped network architecture to enhance long-range dependency modeling in medical image analysis. This enhancement allows the model to simultaneously capture both local detailed features and long-distance dependencies within the images. Unlike the quadratic complexity typically associated with Transformers, our network offers linear scaling with feature size.

Figure 4 illustrates the complete SpineMamba network architecture. Specifically, SpineMamba consists of three main components: an encoder, a decoder, and skip connections. When using Mamba blocks in all encoder blocks without including shape prior modules in the skip connections,

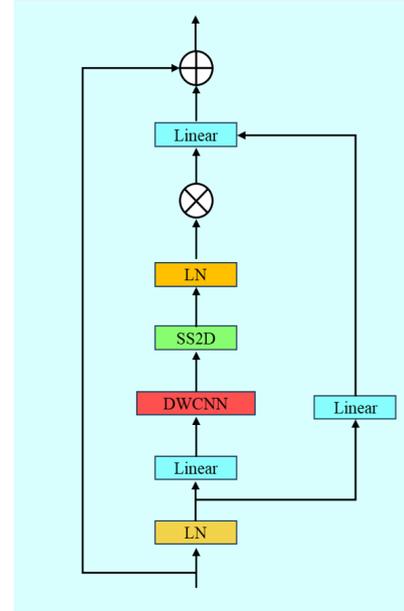

**Fig. 3:** Detailed Structure of the Visual State Space (VSS) Block.

it is denoted as "Enc." When Mamba blocks are only used at the bottleneck, it is denoted as "Bot."

SpineMamba adheres to an encoder-decoder structure, effectively capturing both local features and global context. During preprocessing, SpineMamba adopts the adaptive characteristics of nnU-Net, which automatically determines the number of network blocks across different datasets. The encoder is composed of VSS blocks responsible for feature extraction and downsampling, while the decoder includes VSS blocks and patch expansion operations for upsampling. To enhance responsiveness to spinal anatomical features, shape priors are integrated into the skip connections, maintaining spatial information across different scales. This ensures seamless connection and information flow between the encoder and decoder paths, further improving segmentation performance. Overall, the design of SpineMamba facilitates comprehensive feature learning, capturing complex details and rich semantic context in spinal medical images.

### 3.2. State Space Sequence Model (SSM)

Advanced State Space Models (SSMs) [18], namely Structured State Space Sequence Models (S4) and Mamba, are systems that map 1-D continuous functions or sequences such as $x(t) \in \mathbb{R}$ to $y(t) \in \mathbb{R}$ through a hidden state $h(t) \in \mathbb{R}^N$. Mathematically, this can be expressed as a linear ordinary differential equation (ODE):

$$h'(t) = Ah(t) + Bx(t) \tag{1}$$

$$y(t) = Ch(t) \tag{2}$$

Where, the state matrix $A \in \mathbb{R}^{N \times N}$ serves as the evolution parameter, and $B \in \mathbb{R}^{N \times 1}$ and $C \in \mathbb{R}^{1 \times N}$ serve as projection parameters, and $x(t) \in \mathbb{R}^N$ denotes the implicit latent state.

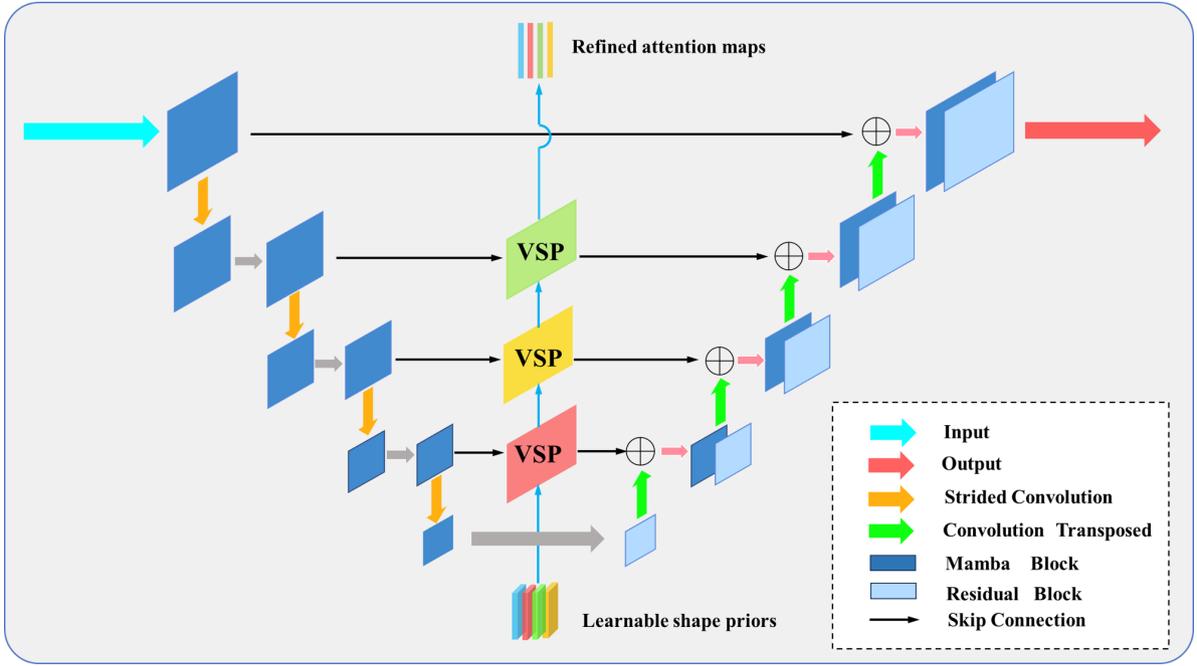

**Fig. 4:** Overview of the SpineMamba (Enc) architecture. The U-Mamba utilizes an encoder-decoder framework, where U-Mamba blocks are employed within the decoder, and residual blocks are also present in the decoder along with skip connections. Here, VSP represents the proposed Shape Prior Module, which is integrated into the skip connections of the U-shaped architecture.

In further improvements, S4 and Mamba discretize this continuous system to make it more suitable for deep learning scenarios. By using the High-Order Polynomial Projection Operator (HIPPO) [52] to construct and initialize the state matrix, they build deep sequence models with rich capabilities and efficient long-range reasoning abilities. Given the input $x(t) \in \mathbb{R}^{L \times D}$, a sampled vector within a signal stream of length $L$, the zero-order hold (ZOH) rule is used to discretize $A$ and $B$ in equation (1) as follows:

$$\overline{A} = \exp(\Delta A) \tag{3}$$

$$\overline{B} = (\Delta A)^{-1}(\exp(\Delta A) - I) \cdot \Delta B \tag{4}$$

where $\Delta \in \mathbb{R}^D$ is the timescale parameter. $\underline{B}, C \in \mathbb{R}^{D \times N}$. The approximation of $\overline{B}$ refined using first-order Taylor series $\overline{B} = \left(e^{\Delta A} - I\right) A^{-1} B \approx (\Delta A)(\Delta A)^{-1} \Delta B = \Delta B$.

Once discretized, the enhanced state space model (SSM)-based formulation can be computationally addressed through two distinct approaches: linear recursion and global convolution, as articulated in Equations 5 and 6, respectively.

$$h'(t) = \overline{A}h(t) + \overline{B}x(t)$$
$$y(t) = Ch(t) \tag{5}$$

$$\overline{K} = (C\overline{B}, C\overline{AB}, \dots, C\overline{A}^{L-1}\overline{B})$$
$$y = x * \overline{K} \tag{6}$$

where $\overline{K} \in \mathcal{R}^L$ represents a structured convolutional kernel, and $L$ denotes the length of the input sequence $x$. Finally, the models produce the output $y$ by applying a global convolution operation with the help of a structured convolutional kernel $\overline{K}$.

### 3.3. Mamba Module Architecture

Recently, the Mamba architecture [19] has significantly optimized the Set Similarity Module (SSM) in discrete data modeling fields such as text and genomic analysis through two key innovations. First, Mamba introduces an input-dependent selection mechanism that, unlike traditional SSMs, is not reliant on fixed time or input, allowing for more effective filtering of information from the input data. This feature is achieved by dynamically adjusting the SSM parameters based on the input data. Second, Mamba has developed a hardware-aware algorithm that scales linearly with sequence length and optimizes operations through loop computation models, enabling Mamba to outperform previous methods on modern hardware.

Therefore, we propose leveraging the linear scaling advantage of Mamba[21] to enhance the representational modeling capability of CNNs. Compared to the quadratic computation complexity of Transformers, Mamba can enhance CNNs' long-range dependency modeling while reducing the required CUDA memory. As shown in Figure 5, each module begins with two consecutive residual blocks, followed by a Mamba block. The residual block contains a plain convolutional layer followed by Instance Normalization (IN) and Leaky ReLU. The image feature tensor, with dimensions $(B, C, H, W, D)$, is flattened and rearranged into dimensions $(B, L, C)$, where $L = H \times W \times D$. After layer normalization, the features flow into a Mamba block, which splits into two parallel paths. In the first path, the features are expanded to $(B, 2L, C)$ via a linear transformation, followed by 1D convolution, SiLU activation, and processing integrated with the SSM layer. In the second path, the features similarly undergo linear expansion and SiLU activation. The features from both paths are then merged. Finally, the features are remapped to the original $(B, L, C)$ format, reshaped, and transposed back to the original $(B, C, H, W, D)$ format. The Mamba architecture achieves significant simplification by integrating the SSM layer with linear layers, demonstrating exceptional computational efficiency during both training and inference stages.

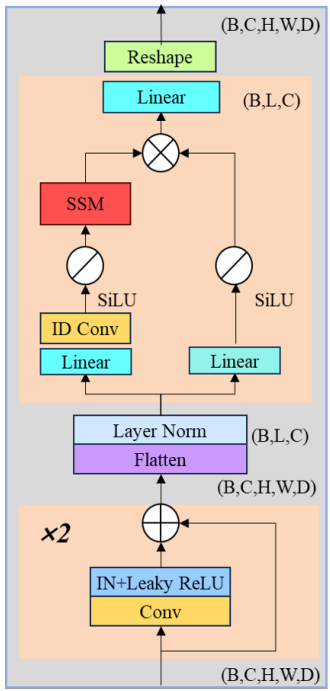

**Fig. 5:** Detailed Structure of the Visual State Space (VSS) Block.

### 3.4. Vertebrae Shape Priors module (VSP)

We propose a learnable Vertebrae Shape Priors module (VSP) that integrates multi-resolution, multi-level features into the long-range dependency modeling process and incorporates this module into a U-shaped neural network architecture. These learnable shape priors are used as inputs to the neural network, combined with the image, to impose anatomical shape constraints on each category of the vertebrae. A set number $N$ of segmentation categories is established as learnable initial templates. By introducing learnable shape priors, specific anatomical shape constraints are applied to each vertebral region, enhancing the representational capability of the U-shaped network. These explicit shape priors can be better interpreted and adjusted as they directly account for the shape and structural information of the vertebral target regions, effectively mitigating noise interference from the background and class similarity interference among vertebrae to a certain extent.

As illustrated in Figure 6, our design seamlessly integrates the Vertebrae Shape Priors (VSP) module into the U-shaped network architecture. The VSP module receives the original image features $F_O$ and the initial vertebrae shape prior $V_0$ as inputs. When these inputs are processed through the feature layers, they yield refined and enhanced features $F_E$ and optimized vertebrae shape priors $V_E$. During this process, the learnable shape priors $S$ are dynamically updated according to the image-label pairs during the training phase. Once training is complete, these learnable shape priors are fixed.

Furthermore, the overall network ultimately predicts a more refined segmentation mask, with the attention map generated by $S$ providing a richer representation of the ground truth region. The shape-prior-based model can be described by the following equation:

$$Y_{\text{predict}} = F\left(X_{\text{predict}}, S\left(X_{\text{train}}, Y_{\text{train}}\right)\right) \qquad (7)$$

where $F$ represents the model's forward propagation during the inference process, and $S$ is the continuous shape prior that constructs the mapping between the image space $I$ and the label space $L$.

Our Vertebrae Shape Priors (VSP) module, through deep integration with a multi-scale U-shaped network architecture, overcomes the traditional limitations of relying solely on deep encoder features. This fusion strategy not only enhances the precise capture of anatomical structures but also significantly improves the accuracy of segmentation tasks. Specifically, the fixed explicit shape priors integrated into the model can generate fine-grained attention maps that accurately identify and locate key regions within the image while effectively suppressing background noise. Notably, even when faced with partially inaccurate ground truth annotations, our learnable shape prior $S$ has demonstrated stability and robustness, further validating the effectiveness of our approach.

Our model adopts an innovative Mamba module to replace traditional self-attention and cross-attention modules, a transformation that significantly optimizes computational efficiency. The Mamba module is highly efficient because it fundamentally avoids the quadratic computational complexity required by traditional attention mechanisms, effectively reducing the consumption of computational resources while preventing issues like numerical instability and gradient explosion due to excessively long token sequences. This design is particularly advantageous when processing large-scale data in a CUDA environment. Although reducing patch size can alleviate the computational burden to some extent, it compromises image resolution, potentially leading to a decline in model performance. In contrast, our Mamba module effectively balances computational efficiency and model performance without sacrificing image resolution. Additionally, our learnable shape priors, supported by convolutional encoder features, generate detailed shape information feature maps. These features are then efficiently processed by the state-space model, significantly enhancing the model's understanding of global context. This deeply integrated design not only maintains precise capture of anatomical structures but also substantially improves segmentation task accuracy.

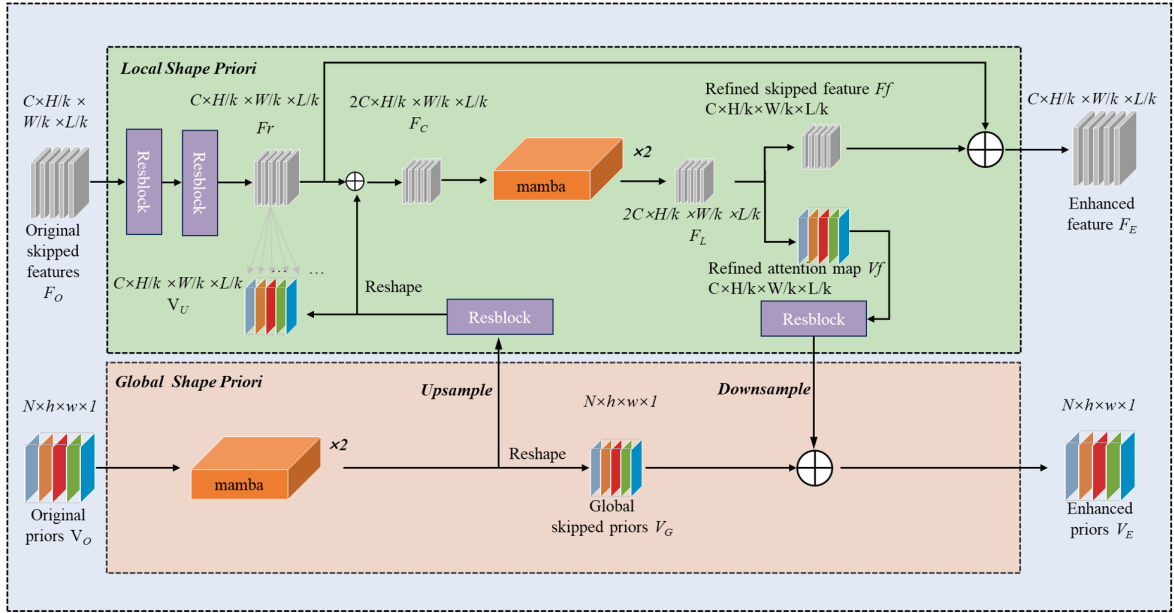

**Fig. 6:** The proposed Vertebrae Shape Priors (VSP) module architecture.

The overall design of the VSP module, as shown in Figure 6, is broadly divided into two main components: global shape priors and local shape priors. In the following sections, we will provide a more in-depth explanation and detailed description of these two modules.

### 3.4.1. Global Shape Prior Module

Considering that channel information can provide rich shape information for the ground truth region, the global shape prior module generates shape priors that facilitate interaction between $N$ channels. The size of the shape priors is preset to $N\times$ spatial dimensions, where $N$ represents the number of different vertebral regions, and the spatial dimensions are related to the patch size. In our designed global shape prior module, we incorporate the integrated modules from Mamba. The input feature is derived from the initial shape prior $V_0$, where the spatial dimensions of $V_0$ are set to $h \times w \times l$, representing a $\frac{1}{16}$ ratio of the patch size $H \times W \times L$. Initially, the features are extracted through two stacked Mamba integrated modules, while maintaining high consistency in the output feature dimensions.

The features extracted from $V_0$ by the Mamba blocks then produce two branches. One branch undergoes an up-sampling process and then interacts within the local shape prior module, while the other branch, after a reshape operation, becomes the global shape prior $V_G$ with dimensions $N \times h \times w \times l$. This global shape prior $V_G$ is then fused with the downsampled shape prior $V_{\text{Re}}$ (originating from the local shape prior module) to obtain the enhanced vertebrae shape prior $V_E$. In this way, the two networks can effectively interact, thereby modeling long-range contextual dependencies and acquiring rich texture information related to the global regions.

$$V_G = \text{Reshape}\left(\text{Mam}(V_0)\right) \tag{8}$$

Here, Mam denotes the processing through two Mamba modules, and $V_0 = V_G = N \times h \times w \times l$. Subsequently, $V_{\text{Re}}$ represents the refined shape prior obtained after splitting. The channel is transformed from $C$ to $N$ using a Resblock, and after downsampling, it is concatenated with the global shape prior $V_G$ to obtain the final enhanced shape prior $V_E$. This allows the network to effectively model long-range contextual dependencies and further acquire rich texture information related to the global regions.

$$V_E = \text{Downsample} \cdot \text{Res}(V_{\text{Re}}) + V_G \tag{9}$$

Here, $\text{Res}(V_{\text{Re}})$ denotes the Resblock operation on $V_{\text{Re}}$, and Downsample represents the downsampling operation.

### 3.4.2. Local Shape Priori Module

The global shape prior module provides the model with rich global contextual information about the spine. However, a single input-output process may lead to the neglect of detailed features of the spine's shape and contours. To address this issue and to introduce appropriate inductive biases into the model for capturing local visual structures and precisely locating objects at different scales, we have meticulously designed a local shape prior module. The inclusion of this module aims to enhance the model's sensitivity to the detailed features of the spine while maintaining a comprehensive understanding of global morphology, ultimately achieving more refined and accurate segmentation results.

Our design integrates the localization and discriminative capabilities of convolution-based regions with the Mamba block's advantages in maintaining linear computational complexity while capturing long-range interactions, as supported by the state-space model (SSM).

As shown in Figure 6, to enable fusion and interaction between the original features $F_O$ from the backbone and the $V_0$ shape priors of different sizes, we first upsample and reshape the original shape prior $V_0$ after processing through two Mamba blocks to obtain $V_U$. Meanwhile, the original feature $F_O$ undergoes two Resblock modules to obtain the feature $F_r$. This ensures that $V_U$ has the same dimensions as $F_r$. Here, $F_O$, $F_r$, and $V_U$ will all have dimensions $C \times H/k \times W/k \times L/k$ (with $k = 2, 4, 8$), where $C$ represents the number of feature channels. Then, $F_r$ and $V_U$ are sequentially concatenated to fuse into $F_C$.

$$V_U = \text{Re} \cdot (\text{Upsample} \cdot \text{Res} * \text{Mam}(V_0)) \quad (10)$$

$$F_C = V_U + F_r \quad (11)$$

Here, $F_C$ has dimensions $2C \times H/k \times W/k \times L/k$ (with $k = 2, 4, 8$). Subsequently, the fused feature $F_C$ undergoes two Mamba modules for refinement, resulting in $F_L$ with the same dimensions.

Next, $F_L$ is sequentially split into $F_f$ and $V_f$. Here, $V_f$ represents the refined vertebral shape prior, which is a matrix of size $C \times H/k \times W/k \times L/k$ (with $k = 2, 4, 8$), and has the characteristic of modeling local visual structures (edges or corners). $F_f$ represents the refined skip features, also a matrix of size $C \times H/k \times W/k \times L/k$ (with $k = 2, 4, 8$), representing the feature mapping relationship between the original feature channels and the original shape prior channels. The enhanced skip feature $F_E$ is the result of concatenating and fusing $F_r$ and $F_f$, which is the final output of the local shape prior module. $F_E$ possesses more precise shape features and rich global textures.

$$F_f = \text{Sep} * F_L, \quad F_L = \text{Mam}(F_C) \quad (12)$$

$$F_E = F_f + F_r \quad (13)$$

Here, Sep $*$ indicates the sequential splitting operation on $F_L$. The module as a whole evaluates the relationship between the $C$ channel feature $F_O$ and the $N$ channel shape prior $V_0$ through the interrelation of $F_L$, $F_f$, and $V_f$.

# 4. Experiment setup

## 4.1. Datasets introduction

We evaluated the performance and scalability of our proposed model on two datasets with different image sizes,

segmentation targets, and modalities: a publicly available CT spinal imaging dataset and a private MR spinal imaging dataset.

**CT dataset:** We used the 41 subjects from the 3Dspine1K data located in the "verse" folder of the large-scale spine CT dataset, CTSpine1K [58]. The in-plane resolutions of the images are all $512 \times 512$, and the number of slices ranges from 315 to 1214 . Due to the absence of annotations for many vertebrae segments in the public dataset, additional processing was required. For more details, please refer to the dataset CTSpine1K.

**MR dataset:** A total of 215 T2-weighted MR volumetric images of the spine were provided by the China Society of Image and Graphics Challenge on Automated Multi-class Segmentation of Spinal Structures on Volumetric MR Images (MRSpineSeg Challenge, https://www.spinesegmentationchallenge.com). We used a 5-fold cross-validation to divide the training and test datasets. Specifically, the dataset was randomly split into 5 folds, each consisting of 43 subjects. Four folds were used for training the model, and the remaining fold was used for testing.

Since we focused on the vertebral bodies, we re-cleaned the labels in this dataset and constructed a 10-class label setup. The vertebrae and intervertebral discs (IVDs) were manually delineated by a junior expert and then corrected by a senior expert using ITK-SNAP [59]. The mean inter-rater intra-class correlation coefficient for the segmentation volumes of the 19 spinal structures was 94.75%. The delineated mask corrected by the senior expert was used as the ground truth for spine parsing. Thus, each subject has a T2-weighted MR image and a corresponding mask, with each vertebra or IVD assigned a unique label. The in-plane resolutions range from $512 \times 512$ to $1024 \times 1024$, and the number of slices ranges from 12 to 18. Detailed information about the spinal disorders in this dataset can be found in Pang et al. [6].

## 4.2. Implementation and training protocols

We implemented SpineMamba within the nnU-Net framework. The modular design of nnU-Net allows for unified control over variables such as image preprocessing and data augmentation, and it can automatically configure hyperparameters for different segmentation datasets during training. As shown in Table 1, the patch size, batch size, and network configurations (e.g., the number of resolution states and the number of downsampling operations along different axes) were kept consistent with nnU-Net. SpineMamba was also optimized using stochastic gradient descent with the loss function being the unweighted sum of Dice loss and cross-entropy. This composite loss further enhanced the model's robustness across different datasets.

Table 1
Configurations for MR and CT datasets

| Configurations | Patch Size | Batch Size | Stages | Pooling per Axis |
|---|---|---|---|---|
| MR | (8, 640, 320) | 2 | 6 | (1, 6, 6) |
| CT | (96, 224, 112) | 2 | 6 | (5, 4, 4) |

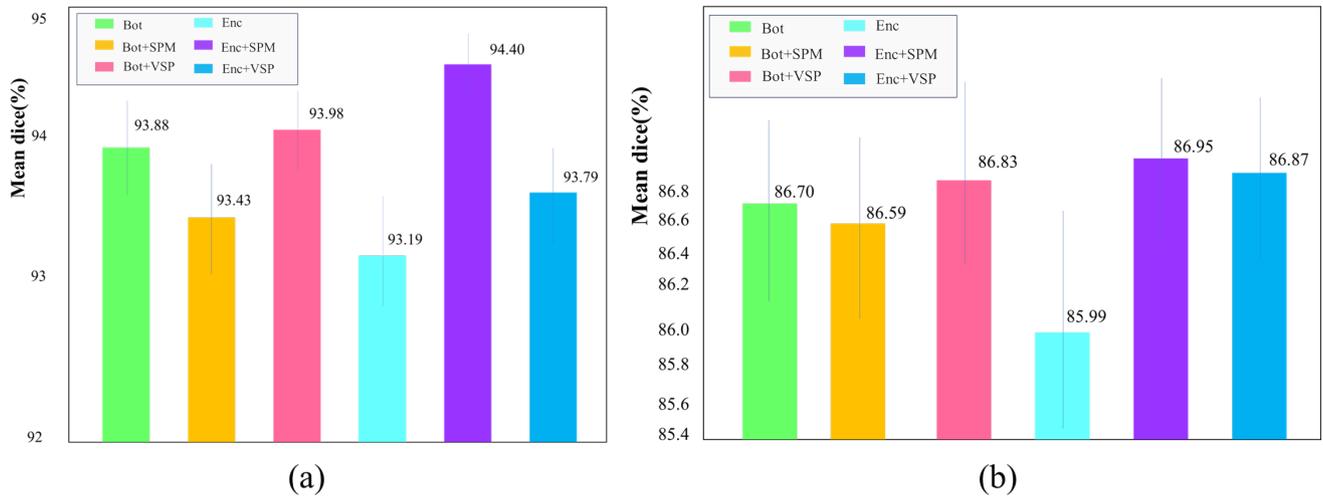

**Fig. 7:** Overall mean Dice (%) of different shape prior modules in two datasets, including two variants of the proposed method. (a) CT dataset (b) MR dataset.

### 4.3. Benchmarking

We conducted comparative experiments using two CNN-based segmentation algorithms, nnU-Net [60] and SegRes-Net [61], and a Transformer-based network, SwinUNETR [48]. In nnU-Net, we utilized the default image preprocessing steps. All networks were integrated into nnU-Net, trained from scratch for 300 epochs on a single NVIDIA A6000 GPU with the same batch size (see Table 2), to ensure architectural consistency. We used the Dice Similarity Coefficient (DSC) as the evaluation metric for segmentation results.

### 4.4. Evaluation method

During the experiments, we performed training data using five-fold cross-validation and averaged the segmentation results from five iterations. Dice was used as a quantitative measure to assess segmentation performance, with each model's Dice score for individual spinal structures (i.e., Vertebrae or IVDs) representing the mean value obtained from five-fold cross-validation.

## 5. Experimental results

### 5.1. Comparison of shape prior module

In this section, we compare the designed visual shape prior module. Specifically, to fairly compare the visual shape prior module, we integrated SPM [46] and our designed VSP into the same baseline Mamba network architectures—Bot and Enc. Figures 8 and 9 show the segmentation results of our designed VSP learnable visual shape prior module method on two datasets. From the visualized results, it can be observed that in most VB and IVD segmentations, both Bot and Enc models based on the Mamba network architecture achieved excellent segmentation performance. As shown in Tables 2 and 3, as expected, the overall average Dice score is higher for the model with our proposed VSP shape prior module added compared to the model with SPM added on the same baseline models (Bot or Enc).

### 5.2. Effectiveness of VSP

Compared to the SPM, our designed shape prior module (VSP) outperformed SPM in most vertebrae body (VB) and intervertebral disc (IVD) segmentations (as shown in Tables 2 and 3). Specifically, the SpineMamba (Bot) + Ours configuration achieved the highest overall mean and the lowest standard deviation, indicating its effectiveness in both segmentation performance and model stability (as illustrated in Figure 7). We observed that in the first row of Figure 8, the overall segmentation results closely match the original labels. However, the Bot+SPM and Enc+SPM models misclassified T9 as background. Additionally, the Enc+SPM model made semantic errors in recognizing L5. In the second row, the Bot-based models performed well in segmentation, but the Enc baseline model produced poor segmentation results for T9. The Enc+SPM model subsequently filled in the main part of the T9 vertebrae body but misclassified the spinous process. In contrast, the Enc+VSP not only optimized the T9 vertebrae body but also retained the baseline model's good segmentation of the T9 spinous process. Additionally, in the third row, the Bot+SPM model misclassified the spinous process pixels of T9 as T10, while the Bot+VSP avoided this issue, maintaining high consistency with the original label and even clearly segmenting the L5 region. These observations validate the effectiveness of our designed shape prior module over the SPM strategy.

Similar results can be observed in the MR dataset. Compared to the Bot baseline, in the second row of Figure 9, the Bot+VSP successfully segmented the coccyx and ensured semantic coherence. In the third row, the Bot+SPM model still produced unclear segmentation for L2, even performing worse than the Bot baseline, but the model with VSP successfully segmented the L2 region. Moreover, due to severe semantic confusion of T10, T11, T12, and L1 in the Enc baseline, the Enc+SPM model failed to correct these errors and included L3 incorrectly. In contrast, the Enc+VSP model corrected the semantic errors in

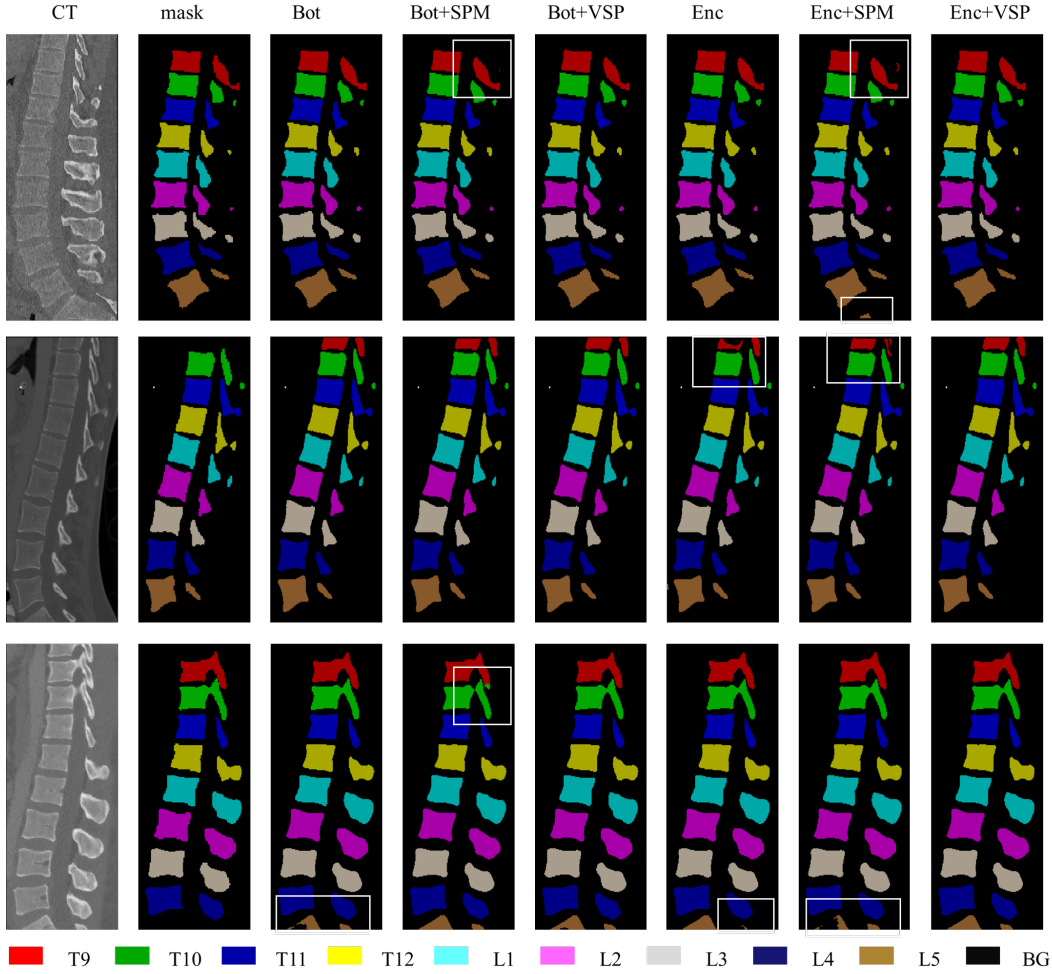

**Fig. 8:** Visualization results of different shape prior modules in CT dataset, including two variants of the proposed method. Each row is a sagittal slice of a subject, and 'BG' represents background.

**Table 2**

Results summary of 3D vertebra segmentation in CT datasets. SpineMamba has the highest mean DSC in most divisions of individual vertebrae.

| Methods | T9 | T10 | T11 | T12 | L1 | L2 | L3 | L4 | L5 | Mean |
|---|---|---|---|---|---|---|---|---|---|---|
| SpineMamba(Bot) | 0.8899±0.14 | 0.9177±0.10 | 0.9252±0.09 | 0.9561±0.02 | 0.9584±0.01 | 0.9594±0.02 | 0.9553±0.03 | 0.9491±0.04 | 0.9372±0.06 | 0.9388±0.06 |
| SpineMamba(Enc) | 0.8976±0.12 | 0.9153±0.11 | 0.9220±0.10 | 0.9531±0.03 | 0.9447±0.05 | 0.9490±0.04 | 0.9530±0.03 | 0.9461±0.05 | 0.9279±0.08 | 0.9343±0.07 |
| SpineMamba(Bot)+SPM | 0.8961±0.13 | 0.9096±0.12 | 0.9246±0.09 | 0.9440±0.06 | 0.9561±0.03 | 0.9630±0.01 | 0.9653±0.01 | 0.9654±0.01 | 0.9351±0.02 | 0.9398±0.05 |
| SpineMamba(Enc)+SPM | 0.8818±0.16 | 0.8980±0.15 | 0.9084±0.14 | 0.9586±0.02 | 0.9591±0.02 | 0.9571±0.02 | 0.9511±0.03 | 0.9435±0.04 | 0.9293±0.07 | 0.9319±0.07 |
| SpineMamba(Bot)+VSP | **0.8995±0.13** | **0.9199±0.11** | **0.9290±0.09** | **0.9573±0.02** | 0.9580±0.02 | 0.9578±0.02 | 0.9604±0.02 | **0.9625±0.01** | 0.9521±0.03 | **0.9440±0.04** |
| SpineMamba(Enc)+VSP | 0.8779±0.16 | 0.8963±0.15 | 0.9067±0.15 | 0.9561±0.03 | **0.9606±0.02** | **0.9649±0.01** | **0.9665±0.01** | 0.9590±0.02 | **0.9534±0.07** | 0.9379±0.06 |

these vertebrae regions, resulting in the best visual results among the models and closely matching the original label (similar observations can be confirmed in the third row of Figure 9). Additionally, using the proposed VSP, the Dice scores for T9, T10, T11 in Table 2 and L1, L3 in Table 3 increased significantly, indicating that the proposed method can improve segmentation performance for small vertebrate bodies (VBs) with class imbalance. The data in the tables also demonstrate significant differences in segmentation results between adding VSP and adding SPM under conditions of limited labeled samples. VSP provided better detail, such as in the segmentation of T9 in the first row of Figure 8 and the segmentation of T12, L1, and L2 in the third row of Figure 9. The model with SPM produced semantic misclassification in the background or semantic confusion between vertebral regions, indicating its failure to correctly constrain the original model with spinal prior knowledge. This result further shows that the learnable shape priors based on Mamba have an advantage over transformer-based methods in learning spinal feature representation. This aids the model in capturing detailed boundary information through the shape prior module. These findings suggest that modules trained with shape prior knowledge play a crucial role in achieving fine-grained spinal segmentation and boundary optimization. Overall, the model with VSP closely aligns with the original mask in most visual results,

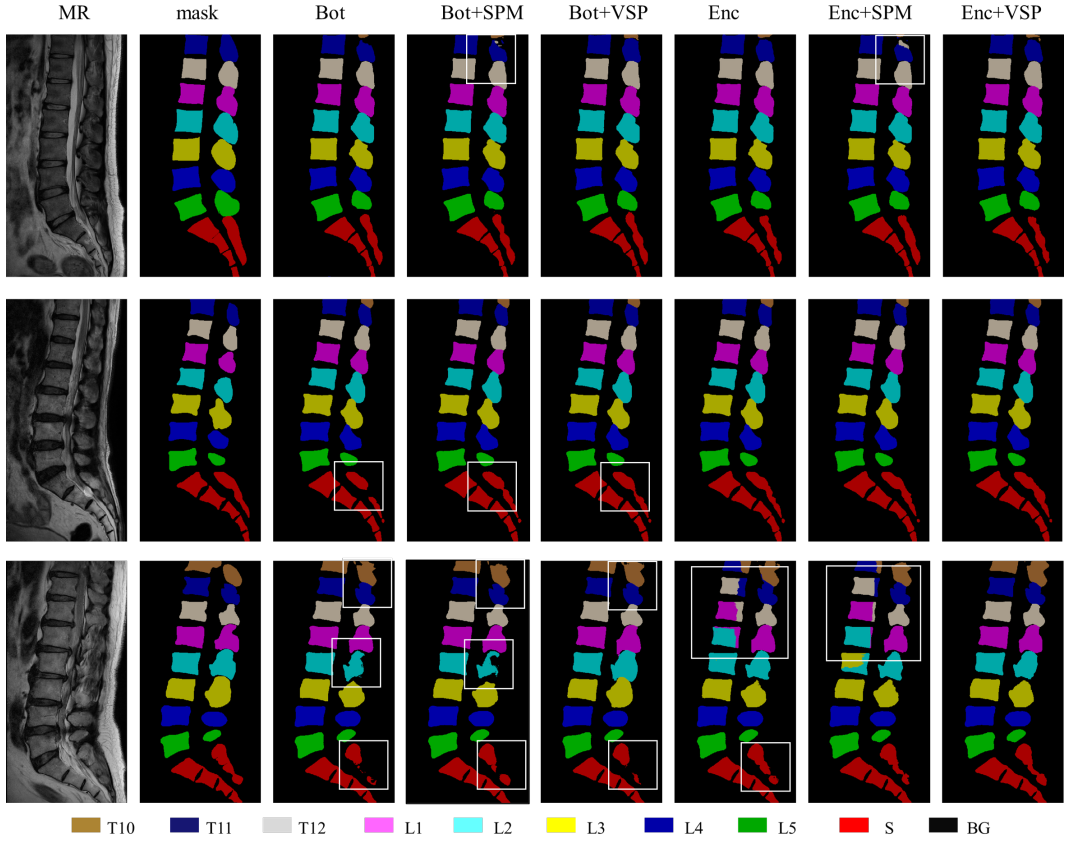

**Fig. 9:** Visualization results of different shape prior modules in MR dataset, including two variants of the proposed method. Each row is a sagittal slice of a subject, and 'BG' represents background.

**Table 3**

Results summary of 3D vertebra segmentation on MR dataset. SpineMamba has the highest mean DSC in most divisions of individual vertebrae.

| Methods | T10 | T11 | T12 | L1 | L2 | L3 | L4 | L5 | S | Mean |
|---|---|---|---|---|---|---|---|---|---|---|
| SpineMamba(bot) | 0.7772±0.18 | 0.8535±0.14 | 0.8737±0.13 | 0.8780±0.13 | 0.8724±0.12 | 0.8792±0.09 | 0.8885±0.06 | 0.8902±0.06 | 0.8901±0.02 | 0.8670±0.10 |
| SpineMamba(Enc) | 0.7610±0.18 | 0.8501±0.15 | 0.8712±0.14 | 0.8786±0.13 | 0.8751±0.11 | 0.8838±0.08 | 0.8926±0.05 | 0.8925±0.03 | 0.8883±0.02 | 0.8659±0.10 |
| SpineMamba(bot)+SPM | 0.7815±0.17 | 0.8525±0.14 | 0.8727±0.13 | 0.8795±0.13 | 0.8742±0.11 | 0.8831±0.10 | 0.8899±0.07 | 0.8915±0.05 | 0.8896±0.02 | 0.8683±0.10 |
| SpineMamba(Enc)+SPM | 0.7461±0.21 | 0.8392±0.17 | 0.8622±0.16 | 0.8712±0.15 | 0.8677±0.13 | 0.8807±0.10 | 0.8916±0.06 | 0.8918±0.04 | 0.8883±0.02 | 0.8599±0.12 |
| SpineMamba(bot)+VSP | **0.7795±0.18** | **0.8562±0.13** | **0.8772±0.12** | **0.8851±0.11** | 0.8785±0.10 | 0.8837±0.08 | 0.8886±0.06 | 0.8887±0.05 | 0.8881±0.02 | **0.8695±0.10** |
| SpineMamba(Enc)+VSP | 0.7665±0.18 | 0.8511±0.14 | 0.8736±0.13 | 0.8837±0.11 | **0.8796±0.10** | **0.8886±0.06** | **0.8946±0.04** | **0.8926±0.04** | 0.8887±0.02 | 0.8687±0.09 |

confirming that our designed VSP is better suited for precise spinal segmentation compared to SPM.

### 5.3. Comparison with other methods

The visualization results of our method compared with other state-of-the-art approaches are presented in Figure 10 and Figure 11, which further evaluates the performance of the proposed method for spinal segmentation. The results shown in Tables 4 and 5, can be summarized as follows:

1. **Segmentation Performance:** The base models of SpineMamba, namely Bot and Enc, outperform nnU-Net on both datasets, demonstrating superior segmentation performance.

2. **SpineMamba's Bot+VSP Model:** The SpineMamba Bot+VSP achieves the highest average Dice scores across both CT and MR datasets. This reinforces the advantage of SpineMamba over existing methods in the context of 3D segmentation tasks.

3. **Impact of the Shape Prior Module:** The shape prior module, VSP, significantly improves the segmentation performance of both the Bot and Enc base models on the two datasets. In contrast, SPM only contributes to enhancing the performance of the Bot base model.

4. **Segmentation Performance for Specific Vertebrae:** The SpineMamba(bot)+VSP model demonstrates the best performance for the first four vertebrae in both datasets. Conversely, the SpineMamba(bot)+VSP model shows improved performance for the L2, L3, and L5 vertebrae. These results indicate the effectiveness of the method in spinal segmentation.

The visualization results of our method compared with other advanced techniques are shown in Figures 10 and 11.

**Table 4**
Results summary of 3D vertebra segmentation in CT dataset.

| Methods | T9 | T10 | T11 | T12 | L1 | L2 | L3 | L4 | L5 | Mean |
|---|---|---|---|---|---|---|---|---|---|---|
| nnU-Net | 0.8373±0.21 | 0.8920±0.15 | 0.9114±0.12 | 0.9315±0.08 | 0.9409±0.05 | 0.9503±0.03 | 0.9612±0.01 | 0.9591±0.02 | 0.9428±0.04 | 0.9252±0.08 |
| SegResNet | 0.8603±0.17 | 0.8667±0.20 | 0.8860±0.18 | 0.9492±0.04 | 0.9542±0.03 | 0.9604±0.01 | 0.9592±0.02 | 0.9539±0.03 | 0.9371±0.05 | 0.9252±0.08 |
| SwinUNETR | 0.8015±0.23 | 0.7839±0.26 | 0.7754±0.27 | 0.8139±0.19 | 0.8642±0.14 | 0.8897±0.10 | 0.9015±0.11 | 0.8895±0.13 | 0.8563±0.17 | 0.8418±0.18 |
| Light-Mamba | 0.8759±0.16 | 0.8981±0.15 | 0.9043±0.15 | 0.9567±0.02 | 0.9552±0.02 | 0.9524±0.03 | 0.9516±0.04 | 0.9450±0.04 | 0.9271±0.06 | 0.9296±0.07 |
| SpineMamba(Enc) | 0.8899±0.14 | 0.9177±0.10 | 0.9252±0.09 | 0.9584±0.01 | 0.9594±0.02 | 0.9553±0.03 | 0.9491±0.04 | 0.9372±0.06 |  | 0.9388±0.06 |
| SpineMamba(Bot) | 0.8976±0.12 | 0.9153±0.11 | 0.9220±0.10 | 0.9531±0.03 | 0.9447±0.05 | 0.9490±0.04 | 0.9530±0.03 | 0.9461±0.05 | 0.9279±0.08 | 0.9343±0.07 |
| SpineMamba(Bot)+SPM | 0.8961±0.13 | 0.9006±0.12 | 0.9246±0.09 | 0.9440±0.06 | 0.9561±0.03 | 0.9630±0.01 | 0.9653±0.01 | 0.9654±0.01 | 0.9351±0.02 | 0.9398±0.05 |
| SpineMamba(Enc)+SPM | 0.8818±0.16 | 0.8980±0.15 | 0.9084±0.14 | 0.9586±0.02 | 0.9591±0.02 | 0.9571±0.02 | 0.9511±0.03 | 0.9435±0.04 | 0.9293±0.07 | 0.9319±0.07 |
| SpineMamba(Bot)+VSP | **0.8995±0.13** | **0.9199±0.11** | **0.9290±0.09** | **0.9573±0.02** | 0.9580±0.02 | 0.9578±0.02 | 0.9604±0.02 | **0.9625±0.01** | 0.9521±0.07 | **0.9440±0.04** |
| SpineMamba(Enc)+VSP | 0.8779±0.16 | 0.8963±0.15 | 0.9067±0.15 | 0.9561±0.03 | **0.9606±0.02** | **0.9649±0.01** | **0.9665±0.01** | 0.9590±0.02 | **0.9534±0.07** | 0.9379±0.06 |

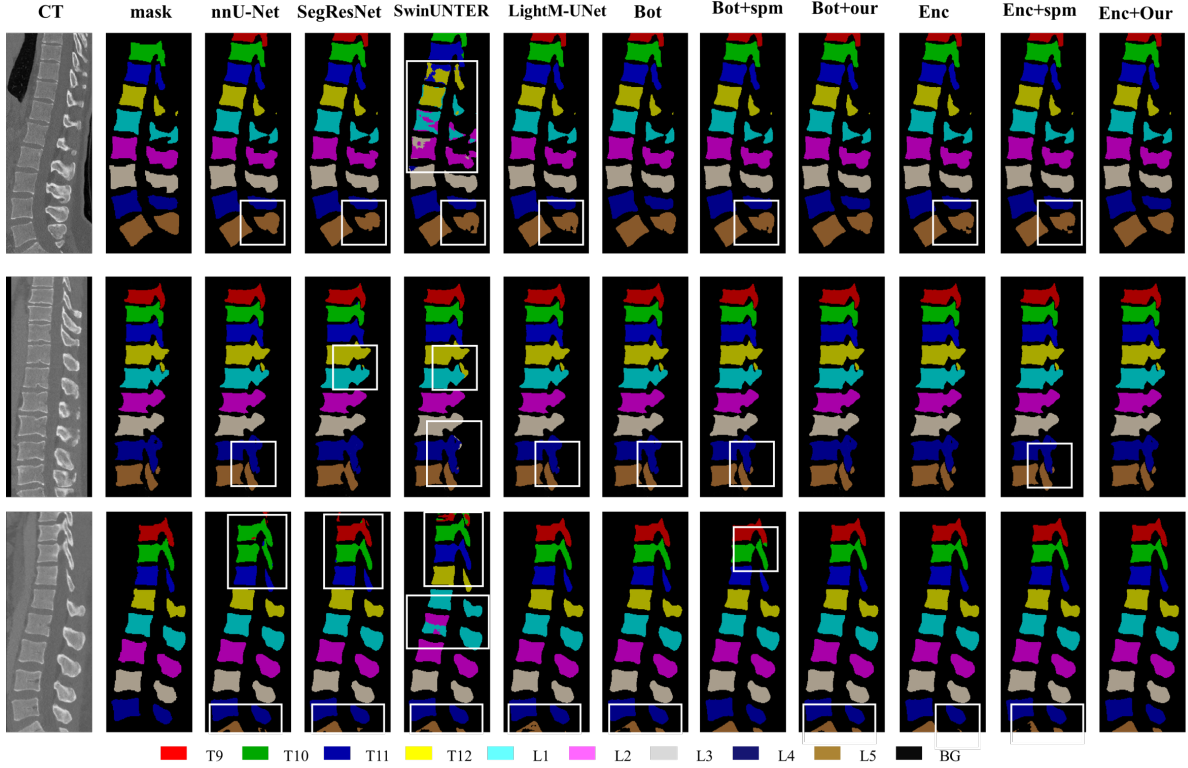

**Fig. 10:** Visualization results of different methods in the CT dataset, where each row is a sagittal slice of a subject, and 'BG' represents background.

1) For the CT dataset, the first row of Figure 10 demonstrates that the proposed Shape Prior Module (SPM) directly enhances the segmentation accuracy of the L5 vertebra, resulting in the most precise segmentation outcome. This indicates that the proposed method can effectively correct suboptimal segmentations caused by morphological errors.

The second row highlights the blurred boundary between the L4 and L5 vertebral arches. The volumetric effects in imaging have led to semantic segmentation omissions of L5 by other advanced methods. In comparison, the segmentation produced by the proposed method more closely aligns with the ground truth mask, suggesting that the method can partially address the challenges posed by volumetric effects in imaging.

**Table 5**
Results summary of 3D vertebra segmentation in MR dataset.

| Methods | T10 | T11 | T12 | L1 | L2 | L3 | L4 | L5 | S | Mean |
|---|---|---|---|---|---|---|---|---|---|---|
| nnU-Net | 0.7733±0.18 | 0.8490±0.15 | 0.8686±0.14 | 0.8743±0.14 | 0.8666±0.13 | 0.8752±0.11 | 0.8814±0.09 | 0.8841±0.07 | 0.8878±0.03 | 0.8622±0.12 |
| SegResNet | 0.6504±0.23 | 0.7910±0.17 | 0.8195±0.17 | 0.8312±0.16 | 0.8336±0.14 | 0.8498±0.10 | 0.8623±0.07 | 0.8626±0.05 | 0.8543±0.04 | 0.8172±0.13 |
| SwinUNETR | 0.6671±0.16 | 0.7750±0.18 | 0.8144±0.16 | 0.8293±0.17 | 0.8354±0.14 | 0.8531±0.11 | 0.8601±0.08 | 0.8563±0.08 | 0.8364±0.04 | 0.8141±0.12 |
| Light-Mamba | 0.7244±0.19 | 0.8069±0.19 | 0.8252±0.19 | 0.8377±0.18 | 0.8424±0.16 | 0.8590±0.12 | 0.8709±0.07 | 0.8687±0.06 | 0.8632±0.03 | 0.8331±0.13 |
| SpineMamba(Bot) | 0.7772±0.18 | 0.8535±0.14 | 0.8737±0.13 | 0.8780±0.13 | 0.8724±0.12 | 0.8702±0.09 | 0.8885±0.06 | 0.8902±0.04 | **0.8901±0.02** | 0.8670±0.10 |
| SpineMamba(Enc) | 0.7610±0.18 | 0.8501±0.15 | 0.8712±0.14 | 0.8786±0.13 | 0.8751±0.11 | 0.8838±0.08 | 0.8926±0.05 | 0.8925±0.03 | 0.8883±0.02 | 0.8659±0.10 |
| SpineMamba(bot)+SPM | 0.7815±0.17 | 0.8525±0.14 | 0.8727±0.13 | 0.8795±0.13 | 0.8742±0.11 | 0.8831±0.10 | 0.8899±0.07 | 0.8915±0.05 | 0.8896±0.02 | 0.8683±0.10 |
| SpineMamba(Enc)+SPM | 0.7461±0.21 | 0.8392±0.17 | 0.8622±0.16 | 0.8712±0.15 | 0.8677±0.13 | 0.8807±0.10 | 0.8916±0.06 | 0.8918±0.04 | 0.8883±0.02 | 0.8599±0.12 |
| SpineMamba(bot)+VSP | **0.7795±0.18** | **0.8562±0.13** | **0.8772±0.12** | **0.8851±0.11** | 0.8785±0.10 | 0.8837±0.10 | 0.8886±0.06 | 0.8887±0.05 | 0.8881±0.02 | **0.8695±0.10** |
| SpineMamba(Enc)+VSP | 0.7665±0.18 | 0.8511±0.14 | 0.8736±0.13 | 0.8837±0.11 | **0.8796±0.10** | **0.8886±0.06** | **0.8946±0.04** | **0.8926±0.04** | 0.8887±0.02 | 0.8687±0.09 |

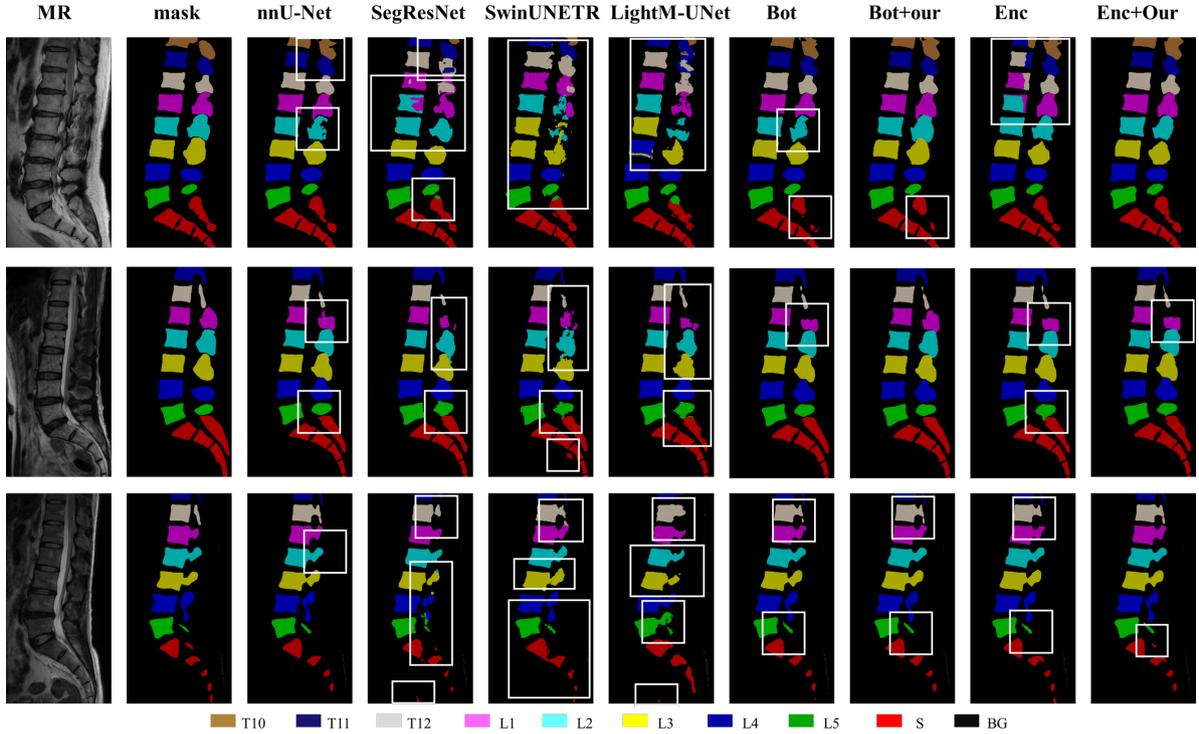

**Fig. 11:** Visualization results of different methods in the MR dataset, where each row is a sagittal slice of a subject, and 'BG' represents background.

The third row, within the white rectangle, further illustrates the semantic confusion some advanced algorithms exhibit between similar vertebrae (T9, T10), including the Bot+SPM model. In contrast, our algorithm demonstrates stable performance across different vertebral regions, with particularly complete segmentation of the L5 vertebra at the bottom.

2) For the MR dataset, the first row of Figure 11 demonstrates that the proposed method (Enc+VSP) achieved accurate segmentation of the sacrum (S) and successfully corrected the segmentation of L2. The second row highlights the discrepancies in the segmentation of L2 and L5 due to volumetric effects. Compared to other methods, our approach more effectively segments blurred bony structures.

The third row illustrates segmentation issues with the vertebrae, particularly within the white rectangle around the T12 and L5 vertebral arches. Some algorithms even misclassified the sacrum (S) within the background. In contrast, our method exhibits fewer extraneous segmentations of nonspinal structures and produces results more closely aligned with the ground truth mask. This indicates the proposed method's capability to address interclass similarity issues, demonstrating the algorithm's stability.

Overall, SpineMamba exhibits significant superiority in these scenarios, reflecting its exceptional ability to capture global context. The Vertebral Shape Prior Module (VSP) not only aids the model in accurately identifying various spinal structures and surpasses other methods that incorporate prior modules in terms of segmentation performance, effectively reducing semantic segmentation errors.

## 6. Discussion

In this study, we introduce an innovative approach named SpineMamba, a state-space model-based method designed for precise spinal segmentation in volumetric CT and MR images. This method ingeniously integrates the Mamba architecture to address the locality limitations of Convolutional Neural Networks (CNNs) and the high computational cost of Transformer architectures, effectively overcoming challenges in long-range dependency modeling. Additionally, the incorporation of the designed Visual Shape Prior (VSP) module optimizes the model, achieving the highest average Dice scores and superior performance in segmenting spinal structures with unbalanced data labels. The method excels in enhancing model generalization and refining segmentation boundary details. Through a comprehensive series of experiments, we demonstrate that SpineMamba consistently outperforms existing CNN and Transformer-based segmentation networks across different modalities and segmentation targets.

This method not only offers a novel approach for automatic and accurate segmentation of complex spinal structures but also holds significant promise for clinical diagnosis and treatment of spinal diseases. Particularly in handling medical objects with specific morphological shapes and anatomical characteristics, SpineMamba exhibits superior segmentation capabilities, significantly reducing the occurrence of outliers. The performance improvement is primarily attributed to the design of the Mamba architecture, which simultaneously extracts multi-scale local features and captures long-range dependencies. Furthermore, addressing the common oversight of morphological priors in most medical segmentation frameworks, our designed learnable shape prior module enhances the model's ability to learn spinal characteristics, providing robust support for precise segmentation.

Moreover, the rational hardware configuration of the SpineMamba model presents significant advantages for its application in clinical practice. The model requires a minimum of only 24GB GPU (Batch size = 1) to run smoothly, a demand well within the capabilities of current medical equipment standards, effectively lowering the barrier to technology deployment. The model's rapid processing capability provides clinicians with swift and precise segmentation results, significantly optimizing clinical workflows, accelerating the diagnostic process, and enhancing clinical decision-making efficiency. Its compatibility with multimodal data and strong generalization ability further broaden its application scenarios across various medical imaging fields, showcasing its potential in clinical practice. In summary, the SpineMamba model, with its exceptional efficiency, accuracy, and broad applicability, significantly contributes to the advancement of clinical spinal imaging segmentation techniques, positioning itself as a powerful tool to drive progress in this domain.

Moreover, there are certain issues with the annotation of the original datasets. For instance, the CT dataset originates from the 'Verse' files in CTspine1k [58]. Notably, the annotations for the coccygeal region are missing, which not only limits our ability to segment this area but also affects the semantic coherence of the entire spinal region's annotations. This omission could potentially lead to adverse effects during the inference process. In our experiments, we observed that even within the same vertebral bone, different segmentation categories could significantly impact the results. Additionally, compared to the CT dataset, the segmentation performance on the MR dataset was notably inferior. We hypothesize the reasons as follows:

Firstly, the CT imaging modality provides a clarity advantage in the bony regions of the spine, reducing the difficulty for segmentation algorithms. Secondly, the MR imaging typically has fewer slices, with only 12-15 images per patient, whereas CT imaging, with smaller inter-slice spacing, can have up to around 200 slices. Therefore, the CT dataset is significantly richer in voxel information and inter-slice semantic coherence compared to the MR dataset. For Transformer architectures, which heavily depend on data quantity, the lack of coherence in MR spinal imaging could negatively impact segmentation algorithms, leading to issues such as semantic confusion and inaccurate identification of specific vertebral regions. In our experiments, even with identical experimental setups, including optimizers, parameters, and the original network, Transformer-based networks trained from scratch underperformed CNN-based networks across all test scenarios. This might also relate to the fact that Transformer architectures are more suited for use in large-scale pretraining and fine-tuning paradigms. To address these challenges, we propose abandoning traditional Transformer-based frameworks and modules, instead leveraging the advantages of state-space models to achieve better performance in spinal segmentation tasks.

In future research, we plan to implement a series of measures to enhance the accuracy and reliability of spinal segmentation. Firstly, we will focus on collecting a broader range of spinal samples to tackle challenges posed by small sample sizes and class imbalance. Additionally, we intend to undertake professional annotation and correction work to address the missing annotations in the coccygeal region of the CTspine1k [58] dataset, ensuring the dataset's completeness and accuracy. As for the model itself, we plan to perform extensive training of SpineMamba on large-scale datasets, aiming to develop a readily deployable segmentation tool while providing pretrained model weights for tasks with limited data. Furthermore, we will explore data augmentation techniques for small datasets and design customized loss functions for objects that require specific shape prior assistance. These measures will further enhance the application effectiveness of the Visual Shape Prior module across various scenarios.

With continuous expansions and optimizations, we are confident that SpineMamba will increasingly play a crucial role in clinical medical image segmentation. It will provide physicians with more accurate and efficient auxiliary tools, assisting them in better clinical diagnosis and treatment. We look forward to making SpineMamba a significant milestone in the field of medical image segmentation through ongoing innovation and improvements, bringing profound impacts to the healthcare industry.

## 7. Conclusion

In this study, we propose an innovative segmentation architecture, SpineMamba, based on the state-space model's Mamba architecture. For the first time, we've designed and introduced a Visual Shape Prior (VSP) module within the Mamba architecture. This module imposes specific spinal feature constraints on the network from a morphological perspective, enabling the model to efficiently learn shape-aware and scale-aware features, thereby enhancing the modality-agnostic representational capabilities of existing 3D spinal segmentation frameworks. Furthermore, this module captures and preserves the natural anatomical features of the spine, further improving the segmentation performance of the model. To validate our approach, we conducted rigorous comparative and ablation studies within a unified framework. This marks the first application of the Mamba architecture as a lightweight strategy for segmenting multimodal clinical datasets of the spine. The proposed SpineMamba achieved remarkable results in the accurate segmentation of spinal structures in volumetric CT and MR images, demonstrating its effectiveness and superiority in clinical dataset segmentation. Through these experiments, we not only showcase the robust performance of the SpineMamba architecture but also laid a solid foundation for future research and applications.

## 8. Declaration of competing interest

The authors declare that they have no known competing financial interests or personal relationships that could have appeared to influence the work reported in this paper.

## 9. Acknowledgments

This work was supported in part by the National Key R&D Project of China (2018YFA0704102, 2018YFA0704104), in part by the Natural Science Foundation of Guangdong Province (No. 2023A1515010673), in part by the Shenzhen Technology Innovation Commission (JSGG20220831110400001, CJGJZD20230724093303007), and in part by the Shenzhen Engineering Laboratory for Diagnosis & Treatment Key Technologies of Interventional Surgical Robots (XMHT20220104009). The authors thank "The Key Laboratory of Biomedical Imaging Science and System, Chinese Academy of Sciences."

## References

[1] Z. Han, B. Wei, A. Mercado, S. Leung, S. Li, Spine-gan: Semantic segmentation of multiple spinal structures, Medical image analysis 50 (2018) 23–35.

[2] C. Chen, D. Belavy, W. Yu, C. Chu, G. Armbrecht, M. Bansmann, D. Felsenberg, G. Zheng, Localization and segmentation of 3d intervertebral discs in mr images by data driven estimation, IEEE transactions on medical imaging 34 (2015) 1719–1729.

[3] X. Baraliakos, J. Listing, M. Rudwaleit, H. Haibel, J. Brandt, J. Sieper, J. Braun, Progression of radiographic damage in patients with ankylosing spondylitis: defining the central role of syndesmophytes, Annals of the rheumatic diseases 66 (2007) 910–915.

[4] Z. Wang, X. Zhen, K. Tay, S. Osman, W. Romano, S. Li, Regression segmentation for m{9}3} spinal images, IEEE transactions on medical imaging 34 (2014) 1640–1648.

[5] Q. Liu, Q. Dou, L. Yu, P. A. Heng, Ms-net: multi-site network for improving prostate segmentation with heterogeneous mri data, IEEE transactions on medical imaging 39 (2020) 2713–2724.

[6] S. Pang, C. Pang, L. Zhao, Y. Chen, Z. Su, Y. Zhou, M. Huang, W. Yang, H. Lu, Q. Feng, Spineparsenet: spine parsing for volumetric mr image by a two-stage segmentation framework with semantic image representation, IEEE Transactions on Medical Imaging 40 (2020) 262–273.

[7] O. Ronneberger, P. Fischer, T. Brox, U-net: Convolutional networks for biomedical image segmentation, in: Medical image computing and computer-assisted intervention–MICCAI 2015: 18th international conference, Munich, Germany, October 5-9, 2015, proceedings, part III 18, Springer, 2015, pp. 234–241.

[8] W. Ji, S. Yu, J. Wu, K. Ma, C. Bian, Q. Bi, J. Li, H. Liu, L. Cheng, Y. Zheng, Learning calibrated medical image segmentation via multi-rater agreement modeling, in: Proceedings of the IEEE/CVF Conference on Computer Vision and Pattern Recognition, 2021, pp. 12341–12351.

[9] J. Chen, Y. Lu, Q. Yu, X. Luo, E. Adeli, Y. Wang, L. Lu, A. L. Yuille, Y. Zhou, Transunet: Transformers make strong encoders for medical image segmentation, arXiv preprint arXiv:2102.04306 (2021).

[10] R. Zhao, Z. Shi, Z. Zou, High-resolution remote sensing image captioning based on structured attention, IEEE Transactions on Geoscience and Remote Sensing 60 (2021) 1–14.

[11] Z. Gu, J. Cheng, H. Fu, K. Zhou, H. Hao, Y. Zhao, T. Zhang, S. Gao, J. Liu, Ce-net: Context encoder network for 2d medical image segmentation, IEEE transactions on medical imaging 38 (2019) 2281–2292.

[12] J. Schlemper, O. Oktay, M. Schaap, M. Heinrich, B. Kainz, B. Glocker, D. Rueckert, Attention gated networks: Learning to leverage salient regions in medical images, Medical image analysis 53 (2019) 197–207.

[13] A. Vaswani, Attention is all you need, Advances in Neural Information Processing Systems (2017).

[14] H. Cao, Y. Wang, J. Chen, D. Jiang, X. Zhang, Q. Tian, M. Wang, Swin-unet: Unet-like pure transformer for medical image segmentation, in: European conference on computer vision, Springer, 2022, pp. 205–218.

[15] Z. Liu, Y. Lin, Y. Cao, H. Hu, Y. Wei, Z. Zhang, S. Lin, B. Guo, Swin transformer: Hierarchical vision transformer using shifted windows, in: Proceedings of the IEEE/CVF international conference on computer vision, 2021, pp. 10012–10022.

[16] A. DOSOVITSKIY, An image is worth 16x16 words: Transformers for image recognition at scale, arXiv preprint arXiv:2010.11929 (2020).

[17] R. Tao, W. Liu, G. Zheng, Spine-transformers: Vertebra labeling and segmentation in arbitrary field-of-view spine cts via 3d transformers, Medical Image Analysis 75 (2022) 102258.

[18] A. Gu, K. Goel, C. Ré, Efficiently modeling long sequences with structured state spaces, arXiv preprint arXiv:2111.00396 (2021).

[19] A. Gu, T. Dao, Mamba: Linear-time sequence modeling with selective state spaces, arXiv preprint arXiv:2312.00752 (2023).

[20] Z. Xing, T. Ye, Y. Yang, G. Liu, L. Zhu, Segmamba: Long-range sequential modeling mamba for 3d medical image segmentation, arXiv preprint arXiv:2401.13560 (2024).

[21] J. Ma, F. Li, B. Wang, U-mamba: Enhancing long-range dependency for biomedical image segmentation, arXiv preprint arXiv:2401.04722 (2024).

[22] J. Ruan, S. Xiang, Vm-unet: Vision mamba unet for medical image segmentation, arXiv preprint arXiv:2402.02491 (2024).

[23] G. Prabhu, Automatic quantification of spinal curvature in scoliotic radiograph using image processing, Journal of medical systems 36 (2012) 1943–1951.

[24] B. Ibragimov, R. Korez, B. Likar, F. Pernuš, L. Xing, T. Vrtovec, Segmentation of pathological structures by landmark-assisted deformable models, IEEE transactions on medical imaging 36 (2017) 1457–1469.

[25] H. Anitha, A. Karunakar, K. Dinesh, Automatic extraction of vertebral endplates from scoliotic radiographs using customized filter, Biomedical Engineering Letters 4 (2014) 158–165.

[26] D. Forsberg, Atlas-based segmentation of the thoracic and lumbar vertebrae, Recent advances in computational methods and clinical applications for spine imaging (2015) 215–220.

[27] F. Fallah, S. S. Walter, F. Bamberg, B. Yang, Simultaneous volumetric segmentation of vertebral bodies and intervertebral discs on fat-water mr images, IEEE journal of biomedical and health informatics 23 (2018) 1692–1701.

[28] B. Glocker, D. Zikic, E. Konukoglu, D. R. Haynor, A. Criminisi, Vertebrae localization in pathological spine ct via dense classification from sparse annotations, in: Medical Image Computing and Computer-Assisted Intervention–MICCAI 2013: 16th International Conference, Nagoya, Japan, September 22-26, 2013, Proceedings, Part II 16, Springer, 2013, pp. 262–270.

[29] B. Glocker, J. Feulner, A. Criminisi, D. R. Haynor, E. Konukoglu, Automatic localization and identification of vertebrae in arbitrary field-of-view ct scans, in: Medical Image Computing and Computer-Assisted Intervention–MICCAI 2012: 15th International Conference, Nice, France, October 1-5, 2012, Proceedings, Part III 15, Springer, 2012, pp. 590–598.

[30] S. Zhao, X. Wu, B. Chen, S. Li, Automatic vertebrae recognition from arbitrary spine mri images by a category-consistent self-calibration detection framework, Medical Image Analysis 67 (2021) 101826.

[31] Y. Zhou, Y. Liu, Q. Chen, G. Gu, X. Sui, Automatic lumbar mri detection and identification based on deep learning, Journal of digital imaging 32 (2019) 513–520.

[32] D. Zhang, B. Chen, S. Li, Sequential conditional reinforcement learning for simultaneous vertebral body detection and segmentation with modeling the spine anatomy, Medical image analysis 67 (2021) 101861.

[33] X. Li, Q. Dou, H. Chen, C.-W. Fu, X. Qi, D. L. Belavỳ, G. Armbrecht, D. Felsenberg, G. Zheng, P.-A. Heng, 3d multi-scale fcn with random modality voxel dropout learning for intervertebral disc localization and segmentation from multi-modality mr images, Medical image analysis 45 (2018) 41–54.

[34] H. Kalinic, Atlas-based image segmentation: A survey, Croatian Scientific Bibliography (2009) 1–7.

[35] Y.-W. Tai, J. Jia, C.-K. Tang, Local color transfer via probabilistic segmentation by expectation-maximization, in: 2005 IEEE Computer Society Conference on Computer Vision and Pattern Recognition (CVPR'05), volume 1, IEEE, 2005, pp. 747–754.

[36] M. Bach Cuadra, V. Duay, J.-P. Thiran, Atlas-based segmentation, Handbook of Biomedical Imaging: Methodologies and Clinical Research (2015) 221–244.

[37] M. Cabezas, A. Oliver, X. Lladó, J. Freixenet, M. B. Cuadra, A review of atlas-based segmentation for magnetic resonance brain images, Computer methods and programs in biomedicine 104 (2011) e158–e177.

[38] T. Cootes, J. Edwards, C. Taylor, Active apperance models. ieee transactions on pattern analysis and machine intelligence, IEEE Transactions on Pattern Analysis and Machine Intelligence 23 (1998) 681685.

[39] I. Matthews, S. Baker, Active appearance models revisited, International journal of computer vision 60 (2004) 135–164.

[40] R. El Jurdi, C. Petitjean, P. Honeine, V. Cheplygina, F. Abdallah, High-level prior-based loss functions for medical image segmentation: A survey, Computer Vision and Image Understanding 210 (2021) 103248.

[41] Y. Xu, Q. Zhang, J. Zhang, D. Tao, Vitae: Vision transformer advanced by exploring intrinsic inductive bias, Advances in neural information processing systems 34 (2021) 28522–28535.

[42] J. Dai, H. Qi, Y. Xiong, Y. Li, G. Zhang, H. Hu, Y. Wei, Deformable convolutional networks, in: Proceedings of the IEEE international conference on computer vision, 2017, pp. 764–773.

[43] R. El Jurdi, C. Petitjean, P. Honeine, F. Abdallah, Bb-unet: U-net with bounding box prior, IEEE Journal of Selected Topics in Signal Processing 14 (2020) 1189–1198.

[44] T.-C. Nguyen, T.-P. Nguyen, G.-H. Diep, A.-H. Tran-Dinh, T. V. Nguyen, M.-T. Tran, Ccbanet: cascading context and balancing attention for polyp segmentation, in: Medical Image Computing and Computer Assisted Intervention–MICCAI 2021: 24th International Conference, Strasbourg, France, September 27–October 1, 2021, Proceedings, Part I 24, Springer, 2021, pp. 633–643.

[45] X. Meng, J. Huang, Z. Li, C. Wang, S. Teng, A. Grau, Dedustgan: Unpaired learning for image dedusting based on retinex with gans, Expert Systems with Applications 243 (2024) 122844.

[46] X. You, J. He, J. Yang, Y. Gu, Learning with explicit shape priors for medical image segmentation, arXiv preprint arXiv:2303.17967 (2023).

[47] J. Chen, Y. Lu, Q. T. Yu, Transformers make strong encoders for medical image segmentation. arxiv 2021, arXiv preprint arXiv:2102.04306 (????).

[48] A. Hatamizadeh, V. Nath, Y. Tang, D. Yang, H. R. Roth, D. Xu, Swin unetr: Swin transformers for semantic segmentation of brain tumors in mri images, in: International MICCAI brainlesion workshop, Springer, 2021, pp. 272–284.

[49] F. Milletari, N. Navab, S.-A. Ahmadi, V-net: Fully convolutional neural networks for volumetric medical image segmentation, in: 2016 fourth international conference on 3D vision (3DV), Ieee, 2016, pp. 565–571.

[50] S. Woo, J. Park, J.-Y. Lee, I. S. Kweon, Cbam: Convolutional block attention module, in: Proceedings of the European conference on computer vision (ECCV), 2018, pp. 3–19.

[51] A. Hatamizadeh, Y. Tang, V. Nath, D. Yang, A. Myronenko, B. Landman, H. R. Roth, D. Xu, Unetr: Transformers for 3d medical image segmentation, in: Proceedings of the IEEE/CVF winter conference on applications of computer vision, 2022, pp. 574–584.

[52] Z. Li, X. Zhou, T. Tong, A two-stage network for segmentation of vertebrae and intervertebral discs: Integration of efficient local-global fusion using 3d transformer and 2d cnn, in: International Conference on Neural Information Processing, Springer, 2023, pp. 467–479.

[53] H. Mehta, A. Gupta, A. Cutkosky, B. Neyshabur, Long range language modeling via gated state spaces, arXiv preprint arXiv:2206.13947 (2022).

[54] J. Wang, W. Zhu, P. Wang, X. Yu, L. Liu, M. Omar, R. Hamid, Selective structured state-spaces for long-form video understanding, in: Proceedings of the IEEE/CVF Conference on Computer Vision and Pattern Recognition, 2023, pp. 6387–6397.

[55] L. Zhu, B. Liao, Q. Zhang, X. Wang, W. Liu, X. Wang, Vision mamba: Efficient visual representation learning with bidirectional state space model, arXiv preprint arXiv:2401.09417 (2024).

[56] Y. Harjoseputro, I. Yuda, K. P. Danukusumo, et al., Mobilenets: Efficient convolutional neural network for identification of protected birds, IJASEIT (International Journal on Advanced Science, Engineering and Information Technology) 10 (2020) 2290–2296.

[57] N. Shazeer, Glu variants improve transformer, arXiv preprint arXiv:2002.05202 (2020).

[58] Y. Deng, C. Wang, Y. Hui, Q. Li, J. Li, S. Luo, M. Sun, Q. Quan, S. Yang, Y. Hao, et al., Ctspine1k: a large-scale dataset for spinal vertebrae segmentation in computed tomography, arXiv preprint arXiv:2105.14711 (2021).

[59] P. A. Yushkevich, A. Pashchinskiy, I. Oguz, S. Mohan, J. E. Schmitt, J. M. Stein, D. Zukić, J. Vicory, M. McCormick, N. Yushkevich, et al., User-guided segmentation of multi-modality medical imaging datasets with itk-snap, Neuroinformatics 17 (2019) 83–102.

[60] F. Isensee, P. F. Jaeger, S. A. Kohl, J. Petersen, K. H. Maier-Hein, nnu-net: a self-configuring method for deep learning-based biomedical image segmentation, Nature methods 18 (2021) 203–211.

[61] A. Myronenko, 3d mri brain tumor segmentation using autoencoder regularization, in: Brainlesion: Glioma, Multiple Sclerosis, Stroke and Traumatic Brain Injuries: 4th International Workshop, BrainLes 2018, Held in Conjunction with MICCAI 2018, Granada, Spain, September 16, 2018, Revised Selected Papers, Part II 4, Springer, 2019, pp. 311–320.